\documentclass[12pt]{article}
\pdfoutput=1
\usepackage{Max}

\usepackage{float}

\title{Distributions of extremal black holes in Calabi-Yau compactifications}
\author{George Hulsey}
\author{Shamit Kachru}
\author{Sungyeon Yang}
\author{Max Zimet}
\affil{Stanford Institute for Theoretical Physics,

Stanford University, Stanford, CA 94305 USA}

\date{}

\begin{document}

\maketitle

\begin{abstract}
We study non-supersymmetric extremal black hole excitations of 4d $\N=2$ supersymmetric string vacua arising from compactification on
Calabi-Yau threefolds.  The values of the (vector multiplet) moduli at the black hole horizon are governed by the attractor mechanism.
This raises natural questions, such as ``what is the distribution of attractor points on moduli space?" and ``how many attractor
black holes are there with horizon area up to a certain size?" 
We employ tools developed by Denef and Douglas \cite{douglas:distributions} to answer these questions.
\end{abstract}

\newpage
\tableofcontents
\hypersetup{linkcolor=blue}

\section{Introduction}

The attractor mechanism \cite{kallosh:attractor,strominger:attractor,kallosh:attractor2} is a ubiquitous phenomenon in gravitational theories with moduli whereby the moduli take on values at the horizon of an extremal\footnote{The attractor mechanism was recently argued to hold in a spatially averaged sense even for non-extremal black holes \cite{goldstein:hotAttr}. Implications of the attractor mechanism for near-extremal black holes were studied in  \cite{larsen:nAttractor}.} (zero temperature) black hole that are independent\footnote{There are two caveats to this that we will discuss below. First, attractor moduli might depend on a choice of `area code.' Second, it is possible for only some moduli to be attracted.} of their values at infinity. It can be heuristically motivated (see, e.g., \cite{trivedi:nonSUSY}) by noting that the entropy of such a black hole -- being the logarithm of an integer -- should vary discretely, while the moduli vary continuously. Alternatively, it can be understood geometrically from the fact that extremal black holes have an infinite throat so that the physical distance to the horizon is infinite, and variations of the moduli over this distance wash away any dependence on the moduli at infinity \cite{kallosh:nonBPS}. Ultimately it follows from an analysis of the solutions of the equations of motion \cite{kallosh:attractor,strominger:attractor,kallosh:attractor2,sen:entropyFunction,kallosh:nonBPS,goldstein:nonSUSY,trivedi:nonSUSY}.

The discussion thus far has made no reference to supersymmetry, but of course the latter is intimately connected to the attractor mechanism. One reason is that the very existence of moduli is generically unnatural without supersymmetry. Another is that the study of BPS attractors is greatly simplified by the first-order Killing spinor equation. 
However, one can study non-supersymmetric but extremal black holes arising as excitations of supersymmetric vacuum solutions.
This makes it natural to study extremal but non-BPS attractors in supergravity.

A particularly natural setting for such a study is in the context of string theory \cite{trivedi:nonSUSYstr}. One important reason is that the attractor mechanism implies a sort of non-renormalization theorem which, under certain conditions, allows one to compute the entropy of a weakly-coupled collection of microscopic ingredients and reliably continue the result to strong coupling where these ingredients form a black hole \cite{legalDocument}. This allows one to generalize the fantastic success of string theory in counting microstates of BPS black holes \cite{stromingerVafa} to the extremal but non-BPS case. Another good reason is that one can hope to generalize the observations of \cite{moore:arithmeticAttractors} relating arithmetic, geometry, and attractors in string theory, those of \cite{denef:attractor,denef:slag,aspinwall:unstableSlag} which pertained to geometric aspects of attractor flows, and those of \cite{sk:specialCycle,sk:cycleAuto} on special cycles. Finally, one might reasonably argue that it is especially interesting to study black holes in a theory with a known UV completion.

First questions one might ask are `How are attractor moduli distributed in moduli space?' and `How many attractors exist with entropy below some cutoff $S_*$?' Denef and Douglas \cite{douglas:distributions} attacked these problems in the BPS case, using tools (reviewed in \cite{douglas:algGeom}) that they had developed for studying flux compactifications (and which were recently applied in another context in \cite{mz:slagCounts}).  They concluded that attractors are uniformly distributed in moduli space and that their count grows as a particular power of $S_*$ that depends only on the dimension of the moduli space. In particular, their results agreed with those of \cite{moore:arithmeticAttractors}, which related the growth of BPS attractors on $K3\times T^2$ to the Smith-Minkowski-Siegel mass formula. Indeed, these tools enabled the prediction of a previously-unknown constant.

The purpose of this paper is to generalize this approach to non-BPS attractors. As one might expect from the analogous results of \cite{douglas:distributions,douglas:nonSUSY} for non-supersymmetric flux vacua, our results are less elegant than in the BPS case. However, we can nevertheless make non-trivial progress. In particular, the growth of attractors with $S_*$ proves to be the same as in the BPS case. Furthermore, after restricting our attention to the simple case of one-modulus models (such as the mirror quintic), we can analytically characterize the distribution of attractors in various regions of moduli space. In contrast to the BPS case, we find a non-trivial dependence on the curvature of moduli space. We verify these predictions with numerics.

An outline of the rest of this paper is as follows. In section \ref{sec:review}, we review the attractor mechanism in the context of type IIB Calabi-Yau compactifications. In section \ref{sec:analytics}, we then analytically study the distribution of attractors. We numerically verify our predictions in the case of a one-modulus model in section \ref{sec:numerics}. We conclude in section \ref{sec:conclusion}.

\section{The attractor mechanism in type IIB Calabi-Yau compactifications} \label{sec:review}

Consider a 4d gravity theory with $U(1)$ gauge fields and moduli $\phi^i$ coupled to the gauge fields via axio-dilaton-like couplings:
\be - f_{\alpha\beta}(\phi^i) F_{\mu\nu}^\alpha F^{\beta \mu\nu} - \frac{1}{2}\tilde{f}_{\alpha\beta}(\phi^i) \epsilon^{\mu\nu\rho\sigma} F_{\mu\nu}^\alpha F^\beta_{\rho\sigma} \ .\ee
The authors of \cite{trivedi:nonSUSY} studied ans\"atze parametrized by choices of electric and magnetic charges that determine extremal black holes in such a theory. They found that solutions are associated to local minima of an effective potential, $V_{eff}(\phi^i)$. Denoting the attractor moduli that determine such a minimum by $\phi^i_0$, they showed that the horizon radius and entropy of the black hole are respectively given by
\be r_H^2 = V_{eff}(\phi^i_0) \ ,\quad S = \frac{A}{4} = \pi r_H^2 \ . \label{eq:r} \ee
In general, there will be multiple such minima \cite{moore:arithmeticAttractors,denef:attractor,denef:quinticBasin}. In these cases, we refer to the multiple basins of attraction as `area codes.' In addition, it is possible for the effective potential to be independent of some moduli. In this case, these moduli will not be attracted, and the entropy of an extremal black hole will not depend on them.

We now specialize in two steps. Our first specialization is to $\N=2$ theories with an $n$-complex-dimensional vector multiplet moduli space $\M$. (Hypermultiplet scalars do not appear in the effective potential.) $\M$ is a special K\"ahler manifold; we denote its K\"ahler potential by $K$ and its metric by $g_{i\bar\jmath} = \partial_i \partial_{\bar\jmath} K$. In addition, there is a central charge, $Z$, which is a holomorphic section of a line bundle $\Li\to \M$ whose covariant derivative is $D_i Z \equiv \partial_i Z + (\partial_i K)Z$. That is, under a K\"ahler transformation
\be K\to K - f - \bar f \ ,\ee
where $f$ is a holomorphic function on $\M$, $Z$ and $D_iZ$ transform as
\be Z \to e^f Z \ ,\quad D_i Z \to e^f D_iZ \ . \ee
The effective potential is then \cite{kallosh:critical}
\be V_{eff} = e^K\brackets{ g^{i\bar \jmath} (D_i Z)(D_{\bar\jmath} \bar Z) + |Z|^2 } \ . \label{eq:Veff} \ee
BPS attractors not only minimize this, but also satisfy
\be D_i Z = 0 \ . \label{eq:BPS} \ee
Said another way, they minimize the rescaled central charge \cite{kallosh:attractor2}
\be \Z = e^{K/2} Z \ , \label{eq:Z} \ee
in the sense that
\be D_i \Z \equiv \partial_i \Z + \frac{1}{2} (\partial_i K) \Z \equiv e^{K/2}D_i Z = 0 \ ,\ee
or
\be \partial_i|\Z|^2 = 0 \ .\ee
So, they separately minimize the two terms of \eqref{eq:Veff}. In contrast, extremal attractors need only minimize \eqref{eq:Veff}. Another important difference is that for BPS attractors critical points of $|\Z|^2$ in the interior of $\M$ are automatically local minima \cite{kallosh:critical}, since $\partial_i\partial_{\bar\jmath} \log |\Z|^2 = g_{i\bar\jmath}$ implies that the Hessian of $|\Z|^2$ at a critical point is a positive definite matrix, whereas for non-BPS critical points of $V_{eff}$ this positive definiteness is not automatic. Finally, we note the following relationships between the central charge, entropy, horizon area, and mass of a BPS black hole:
\be S = \frac{A}{4} = \pi |\Z|^2 = \pi M^2 \ .\ee
These follow from \eqref{eq:r}--\eqref{eq:Z} and the BPS bound $M=|\Z|$.

We now further specialize to $\N=2$ theories obtained via compactification of type IIB string theory on a Calabi-Yau threefold $X$. The moduli space of such a theory locally factors\footnote{This can fail globally \cite{vafa:dManifolds,katz:enhance}.} as a product of (complexified) K\"ahler and complex structure moduli spaces, and the vector multiplet moduli $\phi^i$ of interest are the complex structure moduli. The complex dimension of the complex structure moduli space is $n=h^{2,1}$. A black hole corresponds to a D3-brane wrapping a 3-cycle of $X$, and its electric and magnetic charges correspond to the homology class $\tilde\gamma \in H_3(X,\ZZ)$ representing this 3-cycle, or equivalently its Poincar\'e dual $\gamma\in H^3(X,\ZZ)$. We identify $\Li\to \M$ with the Hodge line bundle $H^{3,0}(X)\to \M$; importantly, the holomorphic 3-form $\Omega$ is then a holomorphic section of this line bundle, which is unique up to multiplication by a nowhere-vanishing holomorphic function of the $\phi^i$. The K\"ahler potential is given by
\be e^{-K}=i\int \Omega\wedge\bar\Omega \ . \label{eq:Kahler} \ee
Finally, for a black hole with charge $\gamma\in H^3(X,\ZZ)$, the central charge is
\be Z = \int \gamma \wedge \Omega \ .\ee
Importantly, $D_i\Omega$ comprise a basis for $H^{2,1}(X)$. Noting that
\be D_i Z = \int \gamma\wedge D_i\Omega \ ,\ee
we find that the BPS attractor equation \eqref{eq:BPS} (and its conjugate) states that the projection of $\gamma$ onto $H^{2,1}(X)\oplus H^{1,2}(X)$ vanishes. That is,
\be \gamma\in H^{3,0}(X) \oplus H^{0,3}(X) \ . \label{eq:niceForm} \ee
Writing this as
\be \gamma = 2\Imag(\bar S \Omega) \ , \label{eq:newEqn} \ee
where $S$ is a constant, provides an interesting reformulation of the BPS attractor condition. For, it consists of $b_3/2=h^{2,1}+1$ complex equations for the same number of unknowns (namely $S$ and $\phi^i$), whereas we started with the $h^{2,1}$ equations \eqref{eq:BPS} in $h^{2,1}$ unknowns. However, for the price of this extra complexity, we have eliminated all derivatives from the equations. We can recover the original equations by studying \eqref{eq:newEqn} in the Hodge decomposition basis for $H^3(X,\CC)$ comprised of $\Omega,D_i\Omega$, and their conjugates. However, in some circumstances it might be more convenient to study this equation in another basis, such as an integral symplectic basis. An analogue of the formulation \eqref{eq:niceForm} of the attractor equations for non-BPS black holes was obtained in \cite{kallosh:nonBPS}.\footnote{As an aside, we note that another reformulation was provided in \cite{vafa:nonSUSY}.}

We now introduce an integral symplectic basis $C^I$, $I=1,\ldots,b_3 = 2h^{2,1}+2$ for $H_3(X,\ZZ)$ and the Poincar\'e dual 3-forms $C_I$. These satisfy
\be \int C_I\wedge C_J = \int_{C^I}C_J = \Sigma_{IJ} = \begin{pmatrix}
\bf 0 & \bf 1 \\
-\bf 1 & \bf 0
\end{pmatrix} \ .\ee
Expanded in this basis, a charge vector $\gamma\in H^3(X,\ZZ)$ has integral coefficients, which are the electric and magnetic charges: $\gamma = Q^I C_I$, where
\be \int C_I\wedge \gamma = Q^J \int C_I\wedge C_J = \Sigma_{IJ} Q^J \equiv Q_I \ . \ee
Another important function of this basis is to provide convenient coordinates on $\M$, namely the periods of $\Omega$:
\be \Pi_I = \int C_I\wedge \Omega \ .\ee
The electric periods $\Pi_a$, $a=1,\ldots,h^{2,1}+1$, provide a set of projective coordinates on $\M$. The magnetic periods are then functions of the electric ones.

As in \cite{douglas:distributions}, we will find it convenient to have a third basis for $H^3(X,\CC)$ (besides the integral symplectic basis and the Hodge decomposition basis), which we will call the orthonormal basis. First, we introduce a vielbein $e^i_A$, $A=1,\ldots,h^{2,1}$, which satisfies $g_{i\bar\jmath} e^i_A e^{\bar\jmath}_{\bar B} = \delta_{A\bar B}$. Then, the orthonormal basis is comprised of $e^{K/2}\Omega$, $e^{K/2}D_A\Omega=e^{K/2} e^i_A D_i\Omega$, and their conjugates. It is so named because \eqref{eq:Kahler} and
\be ie^K\int D_i\Omega\wedge (D_j \Omega)^* = - g_{i\bar \jmath} \Rightarrow ie^K\int D_A\Omega\wedge (D_B \Omega)^* = - \delta_{A\bar B} \ee
imply that this basis diagonalizes the Hermitian form $\avg{\omega_1,\omega_2}=i\int \omega_1\wedge \bar\omega_2$, and the diagonal values are $\pm 1$ in this basis.

\section{Distributions of attractors} \label{sec:analytics}

The quantity of interest in this paper is the count $\N(S_*,\R)$ of attractor black holes with entropy at most $S_*$ and whose attractor moduli lie in the region $\R\subset\M$. We study this by mimicking the approach of \cite{douglas:distributions} to the analogous problem for BPS attractors. We write
\be \N(S_*,\R) = \sum_{\rm attractors} w(\phi^i_0) \theta(S_*/\pi-V_{eff}(\phi^i_0)) \ ,\quad w(\phi^i) = \piecewise{1}{\phi^i\in \R}{0}{\phi^i\not\in\R} \ . \ee
Then, we introduce the Laplace transform representation of the theta function:
\be \N(S_*,\R) = \lim_{\epsilon\to 0^+} \frac{1}{2\pi i} \int_{\epsilon-i\infty}^{\epsilon+i\infty} \frac{d\alpha}{\alpha} e^{\alpha S_*/\pi} \N(\alpha,\R) \ ,\quad 
\N(\alpha,\R) = \sum_{\rm attractors} w(\phi^i_0)\, e^{-\alpha V_{eff}(\phi_0^i)} \ .\ee
Now, we input the attractor equations (using the notation $|dz|^2\equiv \frac{i}{2}dz\wedge d\bar z=dx\wedge dy$ for the integration measure for a complex variable $z=x+iy$):
\be \N(\alpha,\R) = \sum_{Q^IC_I\in H^3(X,\ZZ)} \int_\R |d^n\phi|^2 \, e^{-\alpha V_{eff}} \delta^{2n}(dV_{eff}) |\det d^2 V_{eff}| \, \theta(d^2V_{eff}) \ ; \label{eq:N} \ee
the final theta function schematically indicates that we require the Hessian of $V_{eff}$ to be positive definite so that our critical points are minima. The Jacobian cancels the one for the change of variables $\phi\to dV_{eff}$. Strictly speaking, these expressions require $d^2V_{eff}$ to be the Hessian obtained by differentiating with respect to the real and imaginary parts of $\phi^i$. Equivalently, we can use a unitary change of basis (in order to preserve the eigenvalues of $d^2V_{eff}$) in order to write
\be d^2V_{eff} = 2 \twoMatrix{\partial_{\bar\imath} \partial_j V_{eff}}{\partial_i \partial_j V_{eff}}{\partial_{\bar\imath} \partial_{\bar\jmath} V_{eff}}{\partial_i \partial_{\bar\jmath} V_{eff}} = 2 \twoMatrix{D_{\bar\imath}D_j V_{eff}}{D_iD_j V_{eff}}{D_{\bar\imath}D_{\bar\jmath} V_{eff}}{D_i D_{\bar\jmath} V_{eff}} \ .\ee
The second equality holds when $\partial_i V_{eff}=0$; it makes use of $\partial_i V_{eff}=D_i V_{eff}$ and the fact that when $\partial_iV_{eff}=0$ we can replace partial derivatives of $\partial_iV_{eff}$ by covariant derivatives. (Note that these covariant derivatives include appropriate Christoffel symbols, to account for cotangent space indices.) We similarly write
\be \delta^{2n}(dV_{eff}) = \prod_i \delta(D_i V_{eff}) \delta(D_{\bar\imath} V_{eff}) \ .\ee
With this, we have specified all of the ingredients in \eqref{eq:N}.

We now make a continuum approximation, which should be accurate in the regime of interest where $S_*$ is large:
\be \N(\alpha,\R) \approx \int d^{2n+2}Q\, \int_\R |d^n\phi|^2 \, e^{-\alpha V_{eff}} \delta^{2n}(dV_{eff}) |\det d^2 V_{eff}| \, \theta(d^2V_{eff}) \ .\ee
Rescaling\footnote{More precisely, we consider our integrals over $Q^I$ -- which parametrize $\RR^{2n+2}$ -- as integrals over the slice $\bar Q = Q$ of the space $\CC^{2n+2}$. We then rescale $Q\to Q/\sqrt{\alpha}$, so that $V_{eff}\to V_{eff}/\alpha$. Finally, we rotate the integration contours so that we again localize to the slice $\bar Q = Q$. Upon doing so we can again think of $Q^I$ as being real and running from $-\infty$ to $\infty$.

Note that when we first regard $Q$ as a complex variable, we must be careful to continue the integrand so that it depends holomorphically on $Q$. In particular, we must take $Q$ outside of the absolute values defining $V_{eff}$ before regarding it as complex. Otherwise, we would need to introduce another set of integration variables $Q^*$ parametrizing another copy of $\CC^{2n+2}$ and regard the integral as being over the slice $Q^*=\bar Q=Q$. We would then rescale $Q\to Q/\sqrt{\alpha}, \, Q^*\to Q^*/\sqrt{\alpha}$ and then analytically continue.

Finally, we address the presence of the theta and delta functions, which are non-analytic distributions, and which indeed are only defined when $V_{eff}$ is real. (For the theta function, this is obvious; for the delta function, a na\"ive definition for complex $V_{eff}$ would yield $4n$ real conditions for the $2n$ real integration variables $\phi$ to satisfy.) First, we use the homogeneity of these distributions under multiplication of their arguments by a non-negative number to divide these arguments by $V_{eff}$, while simultaneously multiplying the integrand by the appropriate power of $V_{eff}$ to leave the integral invariant. Then, we write the $Q$ integral in spherical coordinates, with radial coordinate $q=|Q|$. The theta and delta functions are now independent of $q$ and are not affected by the rescaling $q\to q/\sqrt{\alpha}$. The analytic continuation of $q$ now is unaffected by the theta and delta functions. Finally, once we are done with analytic continuation, it is again the case that $V_{eff}$ is non-negative, so we can again use the homogeneity of the theta and delta functions to return them to their original forms.} $Q\to Q/\sqrt{\alpha}$ rescales $V_{eff}\to V_{eff}/\alpha$ and allows us to easily determine the scaling with $S_*$:
\be \N(S_*,\R) \approx \frac{(2S_*)^{n+1}}{(n+1)!} \int_\R |d^n\phi|^2 \, \det g(\phi) \, \rho(\phi) \ , \label{eq:NS} \ee
where
\be \rho = \frac{1}{(2\pi)^{n+1} \det g} \int d^{2n+2}Q \, e^{-V_{eff}}\delta^{2n}(dV_{eff})|\det d^2V_{eff}|\,\theta(d^2V_{eff}) \ . \label{eq:density} \ee
(To perform the $\alpha$ integral yielding this answer, one can either close the contour to the left and use the Cauchy integral formula or use repeated integration by parts until the integral reduces to the Laplace transform representation of $\theta(S_*/\pi)=1$.)
We have thus answered one of our questions: the growth with $S_*$ is identical to that found in \cite{douglas:distributions} for the BPS case.

In order to study the distribution of attractors in moduli space, we would now like to simplify $\rho$. To do so, we re-write $\gamma$ in the Hodge decomposition basis:
\be \gamma = e^{K/2}\brackets{ i\bar X \Omega - i {\bar Y}^i D_i \Omega + {\rm c.c.} } \ .\ee
These coefficients are chosen so that
\be Z = \int \gamma\wedge \Omega = i Xe^{K/2} \int \Omega\wedge \bar\Omega = Xe^{-K/2} \Rightarrow X = \Z \ee
and
\be D_i Z = \int \gamma\wedge D_i\Omega = -i Y^{\bar \jmath} e^{K/2} \int D_i \Omega \wedge (D_j \Omega)^* = e^{-K/2} g_{i\bar \jmath} Y^{\bar \jmath} \Rightarrow Y_i = D_i \Z \ . \ee
In terms of these variables, we have
\be V_{eff} = Y_i \bar Y^i + |X|^2 \ .\ee
Derivatives of the effective potential may be simplified via the special geometry identities
\be D_i X = Y_i \ ,\quad D_i \bar X = 0 \ ,\quad D_i Y_j = \F_{ijk} \bar Y^k \ , \quad D_i \bar Y_{\bar \jmath} = g_{i\bar \jmath} \bar X \ , \ee
where
\be \F_{ijk} = ie^K \int \Omega \wedge D_i D_j D_k \Omega = ie^K \int \Omega \wedge \partial_i \partial_j \partial_k \Omega \ .\ee
We have the following derivatives of $V_{eff}$ at a local minimum:
\begin{align}
\partial_j V_{eff} &= D_j V_{eff} = \F_{j k \ell}\bar Y^k \bar Y^\ell + 2 Y_j \bar X = 0 \\
\partial_i \partial_j V_{eff} &= D_i D_j V_{eff} = (D_i \F_{jk\ell}) \bar Y^k \bar Y^\ell + 4 \F_{ijk} \bar X \bar Y^k \\
\partial_{\bar\imath} \partial_j V_{eff} &= D_{\bar\imath} D_j V_{eff} = 2g_{j\bar\imath}|X|^2 + 2 Y_j \bar Y_{\bar\imath} + 2 g^{k\bar m} \F_{jk\ell} \bar\F_{\bar\imath \bar m \bar n}\bar Y^\ell Y^{\bar n} \ .
\end{align}
We will shortly employ these to flesh out the expressions in \eqref{eq:density} more fully.

Finally, we adopt the orthonormal basis, where the Jacobian for the change of variables from $Q^I$ to $(X,\bar X,Y^{\bar A},\bar Y^A)$ is $2^{n+1}$ \cite{douglas:distributions}. So,
\be \rho = \frac{1}{\pi^{n+1}} \int |dX|^2\, |dY^{\bar A}|^2\, e^{-V_{eff}} \delta^{2n}(d_AV_{eff})|\det d^2_A V_{eff}|\,\theta(d^2_AV_{eff}) \ ,\ee
where
\be V_{eff} = Y_A \bar Y^A + |X|^2 \ .\ee
We have lost the factor of $1/\det g$ because the derivatives of the effective potential are now evaluated in the orthonormal frame:
\begin{align}
\delta^{2n}(d_A V_{eff}) &= \prod_A \delta(D_A V_{eff}) \delta(D_{\bar A} V_{eff}) = |\det e_A^i|^{-2} \delta^{2n}(dV_{eff}) = \det g\, \delta^{2n}(dV_{eff}) \\
d^2_A V_{eff} &= 2\twoMatrix{D_{\bar A} D_B V_{eff}}{D_A D_B V_{eff}}{D_{\bar A}D_{\bar B}V_{eff}}{D_A D_{\bar B} V_{eff}} = \twoMatrix{e_B{}^j}{}{}{e_{\bar B}{}^{\bar \jmath}} d^2V_{eff} \twoMatrix{e^{\bar \imath}{}_{\bar A}}{}{}{e^i{}_A} \\
\det d^2_A V_{eff} &= |\det e_A^i|^4 \det d^2 V_{eff} = (\det g)^{-2} \det d^2 V_{eff} \ .
\end{align}
Note that positive-definiteness of $d^2 V_{eff}$ is equivalent to that of $d^2_A V_{eff}$, since for any vector $x$,
\be x^\dagger d_A^2V_{eff} \,x = (x')^\dagger d^2 V_{eff} \,x' \ , \ee
with $x'=\twoMatrix{e^{\bar \imath}{}_{\bar A}}{}{}{e^i{}_A} x$. In the BPS case the delta function sets $Y_A=0$, the theta function is unnecessary, the determinant simply contributes $|X|^{2n}$, and the integrals are easily evaluated to give $\pi n!$. That is,
\be \rho_{BPS} = \frac{n!}{\pi^n} \ .  \label{eq:bps} \ee
So, $\rho$ is independent of moduli, meaning that attractors are uniformly distributed in moduli space \cite{douglas:distributions}. We will soon see that this conclusion is altered in the non-BPS case.

In order to make progress in this more general setting, we now specialize to the case $n=1$, which includes the famous mirror quintic studied in \cite{cogp}, as well as the more general one-parameter manifolds studied in \cite{klemm:oneMod,greene:collapse,doran:oneParam,greene:tunneling}. We write $Y\equiv Y_1$, and similarly $\F\equiv \F_{111}$ and $D\F \equiv D_1\F$ (where components refer to the orthonormal basis). We then have
\begin{align}
\partial V_{eff} &= \F \bar Y^2 + 2 Y \bar X = 0 \\
\partial^2 V_{eff} &= (D\F)\bar Y^2 + 4 \F \bar X \bar Y \\
\bar\partial \partial V_{eff} &= 2|X|^2 + 2 |Y|^2 + 2 |\F|^2 |Y|^2 \ . \label{eq:diag}
\end{align}
To deal with the theta function, we notice that positivity of both eigenvalues of $d^2_A V_{eff}$ is equivalent to positivity of both the trace and the determinant. But, the trace is clearly positive, since \eqref{eq:diag} is, so the theta function simply requires positivity of the determinant.

Setting aside the BPS solutions $Y=0$ to $\partial V_{eff}=0$, we have
\be \delta^2(d_A V_{eff}) = \frac{1}{4|Y|^2} \delta\parens{X + \frac{\bar\F Y^2}{2\bar Y}} \delta\parens{\bar X + \frac{\F \bar Y^2}{2Y}} \ , \ee
which allows us to perform the $X$ integrals. We thus find
\be \rho_{non-BPS} = \frac{1}{8\pi^{n+1}} \int |dY|^2\, e^{-V_{eff}} \frac{\det d^2_A V_{eff}}{|Y|^2} \, \theta(\det d^2_A V_{eff}) \ ,\ee
where
\be \frac{\det d^2_A V_{eff}}{|Y|^2} = \parens{4 + 10 |\F|^2 + \frac{9}{4} |\F|^4 - |D\F|^2 } |Y|^2 + \brackets{ 2 (D\F) \bar\F^2 Y^2 + {\rm c.c.}} \label{eq:det} \ee
and
\be V_{eff} = \parens{1 + \frac{|\F|^2}{4}} |Y|^2 \ .\ee
As promised, we see that the non-BPS attractors are not uniformly distributed in moduli space.

We now specialize further to special loci in moduli space.

\subsection{$D\F\approx 0$} \label{sec:approx}

We first restrict to regions where we can approximate $D\F\approx 0$.\footnote{Note that $|D\F|$ is invariant under both K\"ahler transformations and multiplication of the einbein by a phase. So, $D\F\approx 0$ is a gauge invariant statement.} Then, \eqref{eq:det} is positive, and the theta function is extraneous. So, we are left with rather simple integrals over $Y$ which yield
\be \rho_{non-BPS} \approx \frac{ 4+9|\F|^2 }{2\pi \parens{4+|\F|^2}} \ . \label{eq:nbps} \ee
Furthermore, we can study corrections to this approximation by continuing to neglect the theta function, but retaining the entirety of \eqref{eq:det}, and again performing the integral over $Y$. The result is
\be \rho'_{non-BPS} \approx \rho_{non-BPS} - \frac{2 \, |D\F|^2}{\pi(4+|\F|^2)^2} \ . \label{eq:nbps2} \ee
In the model that we consider in section \ref{sec:numerics}, we have $D\F\approx 0$ near the Landau-Ginzburg point and in the large complex structure locus.

\subsection{Conifold} \label{sec:conifold}

We now study an example of a region where the theta function eliminates all candidate attractors. Our analysis closely follows that of \cite{douglas:distributions}. A natural place to look for such non-trivial behavior is near the conifold singularity common to the moduli spaces of many one-parameter threefolds. (On the other hand, since stringy effects become important at such points, this might be viewed as more of a mathematical exercise, or a proof of principle that the theta function is, in general, important.)

A natural modulus to employ in the conifold region is the period, $v$, of the vanishing cycle. The metric takes the form
\be g_{v\bar{v}} = c \ln \frac{\mu^2}{|v|^2} \equiv c \xi \ , \ee
where $\mu$ is a constant and $c = e^{K_0}/2\pi$, with $K_0 = K(v=0)$. We also have
\be \F = \frac{i}{c^{1/2}\xi^{3/2}v} \equiv \frac{1}{\epsilon}\ ,\ \ D\F = \frac{i(\xi-3)}{\epsilon^2} \ ,\ee
where $\epsilon \to 0,\xi \to \infty$ as we approach the degeneration. We then have
\begin{align}
\frac{\det(d^2_A V_{eff})}{|Y|^2} &= \parens{4 + \frac{10}{|\epsilon|^2} + \frac{9}{4|\epsilon|^4} - \frac{|\xi - 3|^2}{|\epsilon|^4}} |Y|^2 + \frac{4}{|\epsilon|^4} \Real{i(\xi-3) Y^2} \nonumber \\
&=  -\frac{|\xi-3|^2 |Y|^2}{|\epsilon|^4} + \Oo(1/|\epsilon|^2) +  \Oo(|\xi|/|\epsilon|^4) .
\end{align}
Since this is negative, the theta function eliminates all candidate attractors.

As suggested by the referee, it is of interest whether or not the integrated density of \emph{critical points} -- not just local minima -- of $V_{eff}$ near the conifold point is finite or not. To determine this, we study \eqref{eq:NS}, with $\R$ a small disc about $v=0$, but ignore the theta function which imposes positive-definiteness of the Hessian. We find that the density $\rho$ (without the theta function) diverges as $\xi^2$, so the integrand of \eqref{eq:NS} diverges as $\xi^3$, but this is not fast enough to give a divergent integral. Thus, as in the flux vacuum counting problems in \cite{douglas:distributions}, we do not find a non-integrable density of critical points near the conifold point.

It would be interesting to study attractors at conifold loci of multi-parameter models, as the large curvature of the moduli space
may lead to an enhanced density of attractor points (as happens with the somewhat analogous case of flux vacua).

\section{Numerical verification} \label{sec:numerics}

We turn now to numerical verification of these formulae. We focus on the mirror of an octic in $\WW\PP^4_{1,1,1,1,4}$ (corresponding to $k=8$ in \cite{klemm:oneMod}; throughout this section, we utilize the periods and other geometric quantities determined in this reference, as well as the approximations thereto employed in \cite{sk:taxonomy}). As we discussed in \S\ref{sec:conifold}, the conifold is not a good place to look for attractors. Similarly, one does not find attractors in the large complex structure locus, as the contribution to \eqref{eq:NS} is suppressed by $\det g\approx 0$\footnote{Of course, this statement is not coordinate-independent. More invariantly, $\int_{|\phi|>R} |d\phi|\, \det g \sim 1/\log(4R)$ at large $R$; in particular, this integral is finite.} (and, furthermore, the 4d supergravity approximation breaks down). However, we do find attractors near the Landau-Ginzburg point. As we now explain, $D\F\approx 0$ in this region, and so we can compare with \eqref{eq:nbps} and \eqref{eq:nbps2}.

Using the formula $\Gamma^i_{jk} = g^{i\bar\ell} \partial_j g_{k \bar\ell}$ for the Christoffel symbols of a K\"ahler manifold, and defining $\kappa_{\phi\phi\phi} = -i e^{-K} \F_{\phi\phi\phi} = \int \Omega\wedge \partial_\phi^3 \Omega$ (with $\phi$ the standard coordinate on the complex structure moduli space of the mirror octic), we have
\begin{align}
D\F &= (g_{\phi \bar\phi})^{-2} D_\phi \F_{\phi\phi\phi} = i e^{K} (g_{\phi \bar\phi})^{-2} D_\phi \kappa_{\phi\phi\phi} \\
&= i e^{K} (g_{\phi \bar\phi})^{-2} \brackets{ \partial_\phi + 2(\partial_\phi K) - 3 (g_{\phi\bar\phi})^{-1} (\partial_\phi g_{\phi\bar\phi}) } \kappa_{\phi\phi\phi} \\
&\approx 6.73 ~\phi^4 |\phi| + \Oo(|\phi|^9) \ , \label{eq:small}
\end{align}
where the final line employed formulae
from \cite{klemm:oneMod}. \eqref{eq:small} explains why $D\F\approx 0$ is a good approximation near the Landau-Ginzburg point $\phi=0$.

In order to numerically test \eqref{eq:nbps} and \eqref{eq:nbps2}, we scanned over all $(2Q_{max}+1)^4$ charge vectors $Q\in \ZZ^4$ whose components satisfied $|Q_I| \le Q_{max} \equiv 21$ and searched for local minima of the effective potential associated to each of these charges with $|\phi_0|<0.5$. The homogeneity of $S$ as a function of $Q$ makes it clear that this charge cutoff corresponds to an entropy cutoff. (More precisely, we searched for local minima in the fundamental domain $0\le \arg\phi < 2\pi/8$, as well as each of its images under a rotation by a power of $e^{i\pi/4}$, and then translated all attractors to the fundamental domain. We acted with the appropriate symplectic monodromy matrix on the charges as we did so and eliminated any duplicate attractors -- i.e., attractors with the same charge and moduli. This was computationally efficient, as it produced attractors whose charges did not satisfy $|Q_I|\le Q_{max}$, but whose entropies were nevertheless comparable to attractors whose charges did satisfy this constraint.)

On the other hand, our analytic formulae hold under the assumption that the entropy cutoff $S_*$ is large. Fortunately, as we will see, our $Q_{max}$ is large enough that these constraints are simultaneously satisfied in a window of intermediate entropies, and in this range our analytic formulae agree with the data. (As a test of the fact that the $Q_{max}$ cutoff is responsible for the failure of our analytic results at large $S_*$, we confirmed that lowering $Q_{max}$ shrank this range. We note that our choice of $Q_{max}$ was not large enough for our analytic formulae for BPS attractors to agree with the data over a sizeable window of entropies.) We discarded unphysical solutions with vanishing entropy (discussed in \cite{denef:attractor}), which we characterized as having $S<0.1$, although since our formulae do not apply in this regime this is mostly for aesthetic purposes. Finally, in order to separate BPS and non-BPS attractors, we employed the cutoff $|D_\phi \Z| = 0.1$.

\begin{figure}[H]
\begin{center}
\includegraphics[width=.49\textwidth]{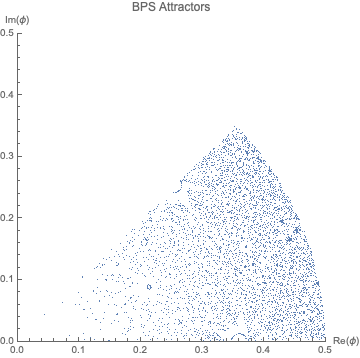}
\includegraphics[width=.49\textwidth]{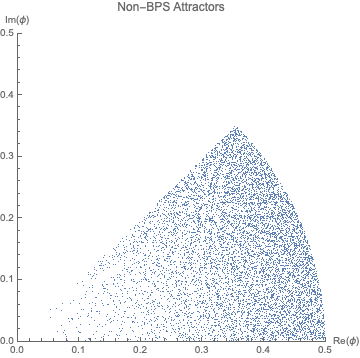}
\caption{Scatter plots of attractor moduli near the Landau-Ginzburg point. Our $\phi$ is the complex structure modulus of \cite{klemm:oneMod}, for which a fundamental domain is $0\le \arg\phi < 2\pi/8$.}\label{fig:scatter}
\end{center}
\end{figure}

\begin{figure}[H]
\begin{center}
\includegraphics[width=.82\textwidth]{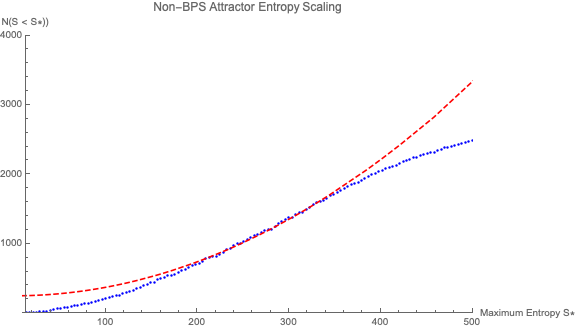}
\caption{Comparison of numerics with the analytic prediction \eqref{eq:NS}, with $\rho$ in that equation replaced by \eqref{eq:nbps}, and where $\R$ is the set $|\phi|<0.5$. We have added a $y$-intercept of $250$ to our theoretical prediction in order to account for the fact that it is only applicable at large $S_*$. Said another way, our theoretical prediction is really for $d\N(S_*)/dS_*$, and we must numerically determine the integration constant. After doing so, we find that the data fits our analytic predictions in an intermediate regime of $S_*$. At larger $S_*$, we have not found all attractors. Because of these limitations at both large and small $S_*$, this figure functions as a weak test of our prediction: it simply demonstrates that at the inflection point of our data, its slope agrees with that of the theoretical result. However, as we explain in the caption of figure \ref{fig:nonBPSarea} it at least correctly identifies the regime in which the data obeys our prediction.}\label{fig:nonBPSscale}
\end{center}
\end{figure}

\begin{figure}[H]
\begin{center}
\includegraphics[width=.82\textwidth]{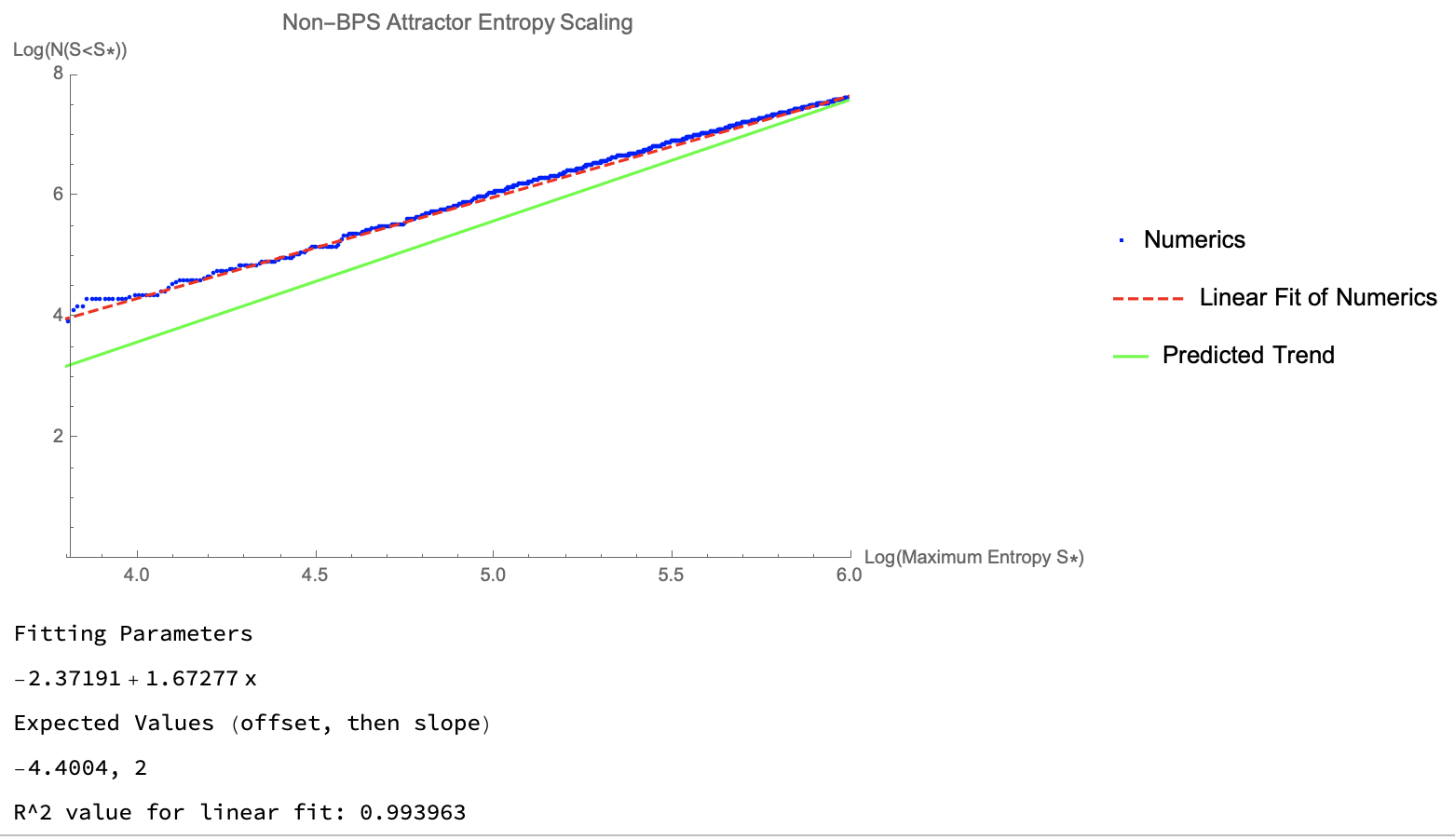}
\caption{Similar to figure \ref{fig:nonBPSscale}, but on a log-log scale. Unlike in figure \ref{fig:nonBPSscale}, we have not added a $y$-intercept to the theoretical prediction. This causes the theoretical prediction to be systematically smaller than the numerical results, thanks to the large number of attractors that were found with small entropy, which are not governed by our large-entropy prediction. However, one can see in this figure that the theoretical prediction is catching up to the numerics, as the total number of attractors begins to overwhelm the number of attractors with small entropies. At the same time, one can discern a trend that is dragging the data down at large $S_*$, which is due to the fact that in this regime we have not found all of the attractors.}\label{fig:nonBPSscaleLogLog}
\end{center}
\end{figure}

\begin{figure}[H]
\begin{center}
\includegraphics[width=.82\textwidth]{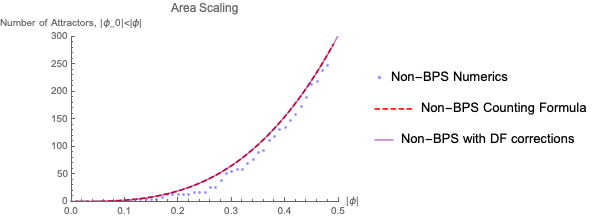}
\caption{Comparison of numerics with the analytic predictions \eqref{eq:nbps} and \eqref{eq:nbps2}. The latter two predictions are indistinguishable. In order to focus on entropies to which our predictions apply, we only include non-BPS attractors with $S_{min}=225< S< S_{max}=275$, where $S_{min}$ and $S_{max}$ are estimated from figure \ref{fig:nonBPSscale}; correspondingly, our theoretical plot is of $\N(S_{max},\R)-\N(S_{min},\R)$, where $\R=\{\phi_0 : |\phi_0| < |\phi|\}$. 
We suspect that inclusion of attractors at the edges of the range $[S_{min},S_{max}]$ is responsible for the evident small systematic errors. As evidence for this, we constructed analogous plots with different choices of $S_{min}$ and $S_{max}$ and observed that the agreement between the numerics and analytic predictions worsened as we increased the size of the interval $[S_{min},S_{max}]$.}\label{fig:nonBPSarea}
\end{center}
\end{figure}

Having said all of this, we now present figures \ref{fig:scatter}, \ref{fig:nonBPSscale}, \ref{fig:nonBPSscaleLogLog}, and \ref{fig:nonBPSarea}.

\section{Conclusion} \label{sec:conclusion}

In this paper, we have studied the distribution of the attractor moduli of non-supersymmetric extremal black holes in the complex structure moduli space of Calabi-Yau threefolds. Growth of attractors with the entropy cutoff $S_*$ was found to be identical to that of the BPS case studied in \cite{douglas:distributions}, owing to the fact that in both cases the effective potential is homogeneous of degree 2 in the charges. However, our results differed in that attractors were not found to be uniformly distributed in moduli space. We illustrated this curvature dependence concretely by focusing on the mirror octic, for which we analytically determined the distribution of attractors in various regions of moduli space. Finally, we tested our predictions near the Landau-Ginzburg point with numerics and found very satisfying agreement.

It would be nice to see some of the geometric and arithmetic connections of BPS attractors generalized to this more physical non-supersymmetric situation. 
For instance, BPS attractors on $K3 \times T^2$ enjoy a close relationship to class groups and class numbers: the U-duality classes of BPS black holes
can be placed in correspondence with binary quadratic forms with discriminant related to the entropy, and the numbers of inequivalent
attractor points at a fixed value of the entropy are governed by the class numbers \cite{moore:arithmeticAttractors,sk:class}.  Finding a similar arithmetic interpretation of
non-BPS attractors on $K3 \times T^2$ is a natural direction to pursue.
We hope that the study of attractors (both BPS and non-BPS) will in some sense provide a natural analogue of the mathematical notion of special cycles for moduli spaces which are not locally symmetric. On the geometric front, the problem of needing to continue weakly-coupled D-branes to a black hole which emerges at stronger coupling is an important obstacle to connecting aspects of Calabi-Yau geometry with gravity solutions, but the successes of string theory in accounting for the entropy of extremal but non-BPS black holes in special cases, as well as those of supergravity in accounting for subtle aspects of BPS spectra in string theory \cite{denef:attractor,denef:correspond}, seem to provide reason for optimism. Furthermore, mathematical tools that might be useful for studying non-BPS extremal cycles have begun to seep into the physics literature \cite{liam:nonHolo}.

\section*{Acknowledgments}

M.Z. thanks Liam McAllister for interesting suggestions. We also thank the anonymous referee for helpful comments. The research of S.K. was supported in part by a Simons Investigator Award and the National Science Foundation under grant number PHY-1720397. 

\newpage
\appendix


\begin{thebibliography}{10}

\bibitem{douglas:distributions}
F.~Denef and M.~R. Douglas, ``{Distributions of flux vacua},'' {\em JHEP}
  {\bfseries 05} (2004) 72,
\href{http://arxiv.org/abs/hep-th/0404116}{{\ttfamily arXiv:hep-th/0404116}}.

\bibitem{kallosh:attractor}
S.~Ferrara, R.~Kallosh, and A.~Strominger, ``{$N=2$ Extremal Black Holes},''
  {\em Phys. Rev.} {\bfseries D52} (1995) 5412--5416,
\href{http://arxiv.org/abs/hep-th/9508072}{{\ttfamily arXiv:hep-th/9508072}}.

\bibitem{strominger:attractor}
A.~Strominger, ``{Macroscopic entropy of $N=2$ extremal black holes},'' {\em
  Phys. Lett.} {\bfseries B383} (1996) 39--43,
\href{http://arxiv.org/abs/hep-th/9602111}{{\ttfamily arXiv:hep-th/9602111}}.

\bibitem{kallosh:attractor2}
S.~Ferrara and R.~Kallosh, ``{Supersymmetry and attractors},'' {\em Phys. Rev.}
  {\bfseries D54} (1996) 1514--1524,
\href{http://arxiv.org/abs/hep-th/9602136}{{\ttfamily arXiv:hep-th/9602136}}.

\bibitem{goldstein:hotAttr}
K.~Goldstein, V.~Jejjala, and S.~Nampuri, ``{Hot Attractors},'' {\em JHEP}
  {\bfseries 01} (2015) 075,
\href{http://arxiv.org/abs/1410.3478}{{\ttfamily arXiv:1410.3478 [hep-th]}}.

\bibitem{larsen:nAttractor}
F.~Larsen, ``{A nAttractor Mechanism for nAdS$_2$/nCFT$_1$ Holography},''
\href{http://arxiv.org/abs/1806.06330}{{\ttfamily arXiv:1806.06330 [hep-th]}}.

\bibitem{trivedi:nonSUSY}
K.~Goldstein, N.~Iizuka, R.~P. Jena, and S.~P. Trivedi, ``{Non-Supersymmetric
  Attractors},'' {\em Phys. Rev.} {\bfseries D72} (2005) 124021,
\href{http://arxiv.org/abs/hep-th/0507096}{{\ttfamily arXiv:hep-th/0507096}}.

\bibitem{kallosh:nonBPS}
R.~Kallosh, N.~Sivanandam, and M.~Soroush, ``{The Non-BPS Black Hole Attractor
  Equation},'' {\em JHEP} {\bfseries 03} (2006) 060,
\href{http://arxiv.org/abs/hep-th/0602005}{{\ttfamily arXiv:hep-th/0602005}}.

\bibitem{sen:entropyFunction}
A.~Sen, ``{Black Hole Entropy Function and the Attractor Mechanism in Higher
  Derivative Gravity},'' {\em JHEP} {\bfseries 09} (2005) 038,
\href{http://arxiv.org/abs/hep-th/0506177}{{\ttfamily arXiv:hep-th/0506177}}.

\bibitem{goldstein:nonSUSY}
D.~Astefanesei, K.~Goldstein, and S.~Mahapatra, ``{Moduli and (un)attractor
  black hole thermodynamics},'' {\em Gen.Rel.Grav.} {\bfseries 40} (2008)
  2069--2105,
\href{http://arxiv.org/abs/hep-th/0611140}{{\ttfamily arXiv:hep-th/0611140}}.

\bibitem{trivedi:nonSUSYstr}
P.~Tripathy and S.~Trivedi, ``{Non-Supersymmetric Attractors in String
  Theory},'' {\em JHEP} {\bfseries 03} (2006) 022,
\href{http://arxiv.org/abs/hep-th/0511117}{{\ttfamily arXiv:hep-th/0511117}}.

\bibitem{legalDocument}
A.~Dabholkar, A.~Sen, and S.~P. Trivedi, ``{Black Hole Microstates and
  Attractor Without Supersymmetry},'' {\em JHEP} {\bfseries 01} (2007) 096,
\href{http://arxiv.org/abs/hep-th/0611143}{{\ttfamily arXiv:hep-th/0611143}}.

\bibitem{stromingerVafa}
A.~Strominger and C.~Vafa, ``{Microscopic origin of the Bekenstein-Hawking
  entropy},'' \href{http://dx.doi.org/10.1016/0370-2693(96)00345-0}{{\em Phys.
  Lett.} {\bfseries B379} (1996) 99--104},
\href{http://arxiv.org/abs/hep-th/9601029}{{\ttfamily arXiv:hep-th/9601029}}.

\bibitem{moore:arithmeticAttractors}
G.~Moore, ``{Arithmetic and Attractors},''
\href{http://arxiv.org/abs/hep-th/9807087}{{\ttfamily arXiv:hep-th/9807087}}.

\bibitem{denef:attractor}
F.~Denef, ``{Supergravity flows and D-brane stability},'' {\em JHEP} {\bfseries
  08} (2000) 050,
\href{http://arxiv.org/abs/hep-th/0005049}{{\ttfamily arXiv:hep-th/0005049}}.

\bibitem{denef:slag}
F.~Denef, ``{(Dis)assembling Special Lagrangians},''
\href{http://arxiv.org/abs/hep-th/0107152}{{\ttfamily arXiv:hep-th/0107152}}.

\bibitem{aspinwall:unstableSlag}
P.~S. Aspinwall, A.~Maloney, and A.~Simons, ``{Black Hole Entropy, Marginal
  Stability and Mirror Symmetry},'' {\em JHEP} {\bfseries 07} (2007) 034,
\href{http://arxiv.org/abs/hep-th/0610033}{{\ttfamily arXiv:hep-th/0610033}}.

\bibitem{sk:specialCycle}
S.~Kachru and A.~Tripathy, ``{BPS jumping loci and special cycles},''
\href{http://arxiv.org/abs/1703.00455}{{\ttfamily arXiv:1703.00455 [hep-th]}}.

\bibitem{sk:cycleAuto}
S.~Kachru and A.~Tripathy, ``{BPS jumping loci are automorphic},''
\href{http://arxiv.org/abs/1706.02706}{{\ttfamily arXiv:1706.02706 [hep-th]}}.

\bibitem{douglas:algGeom}
M.~R. Douglas, ``{Random algebraic geometry, attractors and flux vacua},'' in
  {\em {Encyclopedia of Mathematical Physics}}, J.-P. Fran\c{c}oise, G.~L.
  Naber, and T.~S. Tsun, eds., pp.~323--329.
\newblock Elsevier, 2006.
\newblock
\href{http://arxiv.org/abs/math-ph/0508019}{{\ttfamily arXiv:math-ph/0508019}}.
\newblock

\bibitem{mz:slagCounts}
S.~Kachru, A.~Tripathy, and M.~Zimet, ``{Recounting Special Lagrangian Cycles
  in Twistor Families of K3 Surfaces. Or: How I Learned to Stop Worrying and
  Count BPS States},''
\href{http://arxiv.org/abs/1807.09984}{{\ttfamily arXiv:1807.09984 [hep-th]}}.

\bibitem{douglas:nonSUSY}
F.~Denef and M.~R. Douglas, ``{Distributions of nonsupersymmetric flux
  vacua},'' {\em JHEP} {\bfseries 3} (2005) 61,
\href{http://arxiv.org/abs/hep-th/0411183}{{\ttfamily arXiv:hep-th/0411183}}.

\bibitem{denef:quinticBasin}
F.~Denef, B.~Greene, and M.~Raugas, ``{Split attractor flows and the spectrum
  of BPS D-branes on the Quintic},'' {\em JHEP} {\bfseries 05} (2001) 012,
\href{http://arxiv.org/abs/hep-th/0101135}{{\ttfamily arXiv:hep-th/0101135}}.

\bibitem{kallosh:critical}
S.~Ferrara, G.~W. Gibbons, and R.~Kallosh, ``{Black Holes and Critical Points
  in Moduli Space},'' {\em Nucl. Phys.} {\bfseries B500} (1997) 75,
\href{http://arxiv.org/abs/hep-th/9702103}{{\ttfamily arXiv:hep-th/9702103}}.

\bibitem{vafa:dManifolds}
M.~Bershadsky, V.~Sadov, and C.~Vafa, ``{D-Strings on D-Manifolds},'' {\em
  Nucl. Phys.} {\bfseries B463} (1996) 398--414,
\href{http://arxiv.org/abs/hep-th/9510225}{{\ttfamily arXiv:hep-th/9510225}}.

\bibitem{katz:enhance}
S.~H. Katz, D.~R. Morrison, and M.~R. Plesser, ``{Enhanced gauge symmetry in
  type II string theory},'' {\em Nucl. Phys.} {\bfseries B477} (1996) 105,
\href{http://arxiv.org/abs/hep-th/9601108}{{\ttfamily arXiv:hep-th/9601108}}.

\bibitem{vafa:nonSUSY}
K.~Saraikin and C.~Vafa, ``{Non-supersymmetric Black Holes and Topological
  Strings},'' {\em Class. Quant. Grav.} {\bfseries 25} (2008) 095007,
\href{http://arxiv.org/abs/hep-th/0703214}{{\ttfamily arXiv:hep-th/0703214}}.

\bibitem{cogp}
P.~Candelas, X.~de~la Ossa, P.~Green, and L.~Parkes, ``{A Pair of Calabi-Yau
  Manifolds as an Exactly Soluble Superconformal Theory},'' {\em Nucl. Phys. B}
  {\bfseries 359} (1991) 21.

\bibitem{klemm:oneMod}
A.~Klemm and S.~Theisen, ``{Considerations of One-Modulus Calabi-Yau
  Compactifications: Picard-Fuchs Equations, K\"ahler Potentials and Mirror
  Maps},'' {\em Nucl. Phys.} {\bfseries B389} (1993) 153--180,
\href{http://arxiv.org/abs/hep-th/9205041}{{\ttfamily arXiv:hep-th/9205041}}.

\bibitem{greene:collapse}
B.~R. Greene and C.~I. Lazaroiu, ``{Collapsing D-Branes in Calabi-Yau Moduli
  Space: I},'' {\em Nucl. Phys.} {\bfseries B604} (2001) 181--255,
\href{http://arxiv.org/abs/hep-th/0001025}{{\ttfamily arXiv:hep-th/0001025}}.

\bibitem{doran:oneParam}
C.~F. Doran and J.~W. Morgan, ``{Mirror symmetry and integral variations of
  Hodge structure underlying one-parameter families of Calabi-Yau
  threefolds},''
\href{http://arxiv.org/abs/math/0505272}{{\ttfamily arXiv:math/0505272}}.

\bibitem{greene:tunneling}
P.~Ahlqvist, B.~R. Greene, D.~Kagan, E.~Lim, S.~Sarangi, and I.-S. Yang,
  ``{Conifolds and Tunneling in the String Landscape},'' {\em JHEP} {\bfseries
  03} (2011) 119,
\href{http://arxiv.org/abs/1011.6588}{{\ttfamily arXiv:1011.6588 [hep-th]}}.

\bibitem{sk:taxonomy}
A.~Giryavets, S.~Kachru, and P.~Tripathy, ``{On the Taxonomy of Flux Vacua},''
  {\em JHEP} {\bfseries 08} (2004) 002,
\href{http://arxiv.org/abs/hep-th/0404243}{{\ttfamily arXiv:hep-th/0404243}}.

\bibitem{sk:class}
N.~Benjamin, S.~Kachru, K.~Ono, and L.~Rolen, ``{Black holes and class
  groups},''
\href{http://arxiv.org/abs/1807.00797}{{\ttfamily arXiv:1807.00797 [math.NT]}}.

\bibitem{denef:correspond}
F.~Denef, ``{On the correspondence between D-branes and stationary supergravity
  solutions of type II Calabi-Yau compactifications},''
\href{http://arxiv.org/abs/hep-th/0010222}{{\ttfamily arXiv:hep-th/0010222}}.

\bibitem{liam:nonHolo}
M.~Demirtas, C.~Long, L.~McAllister, and M.~Stillman, ``{The Kreuzer-Skarke
  Axiverse},''
\href{http://arxiv.org/abs/1808.01282}{{\ttfamily arXiv:1808.01282 [hep-th]}}.

\end{thebibliography}

\providecommand{\href}[2]{#2}\begingroup\raggedright\endgroup

\end{document}